\documentclass[twocolumn,letter]{jpsj3}
\usepackage{txfonts}
\def\ve{\varepsilon}

\title{Why the hidden order in URu$_2$Si$_2$ is still hidden --- one simple answer}
\recdate
{\today}

\author{Hisatomo \textsc{Harima}\thanks{E-mail address: hh@kobe-u.ac.jp}$^{1,2}$, 
Kazumasa \textsc{Miyake}$^{2,3}$, Jacques \textsc{Flouquet}$^{2}$
}
\inst{$^{1}$Department of Physics, Graduate School of Science, Kobe University, Nada, Kobe 657-8501, Japan\\
$^{2}$INAC/SPSMS, CEA-Grenoble and 
Universit\'e Joseph Fourier, 17 rue des Martyrs, 38054 Grenoble, France\\
$^{3}$Division of Materials Physics, Department of Materials Engineering Science, Graduated School of Engineering Science,\\
Osaka University, Toyonaka, Osaka 560-8531, Japan
}
\abst{
For more than two decades, the nonmagnetic anomaly observed around 17.5 K in URu$_2$Si$_2$, has been investigated  intensively.
However, any kind of fingerprint for the lattice anomaly has not been observed.
Therefore, the order has been called "the hidden order".
One simple answer to why the hidden order is still hidden is presented from the space group analysis.
The second order phase transition from $I4/mmm$ (No.~139) to $P4_2/mnm$ (No.~136) does not need any kind of lattice distortion in this system, 
and allows the NQR frequency at Ru-site unchanged.
It is compatible with $O_{xy}$-type anti-ferro quadrupole ordering with $\mathbf Q=(0, 0, 1)$.
The characteristics of the hidden order are discussed based on the local $5f^2$ electron picture.
}

\kword{URu$_2$Si$_2$, hidden order, quadrupole ordering, Kondo effect, Nambu-Goldstone mode}
\begin{document}
\maketitle

%\section*{Introduction}

A second order phase transition in the solid states causes change of symmetry;
gauge symmetry, time reversal symmetry, etc... .
On crystal symmetry, 
it usually accompanies lattice distortion to reach lower symmetry.
In exchange for releasing entropy, it becomes a lower energy state in lower temperatures.
The lattice distortion itself needs loss of energy, so another mechanism is necessary to overcome the rising energy in lattice system.
The Jahn-Teller effect is known as such a example.
The degeneracy of the electronic states is lifted, due to distortion, to obtain the lower energy.
Even when the electronic states are not local, such effect is widely discussed in terms of  the Jahn-Teller effect.

The magnitude of such a distortion strongly depends on the mechanism to yield lower energy states.
Lowering the symmetry changes the crystalline electric field (CEF) and the bonding strengths, through the movements of atoms.
The more the lattice is distorted, the much lower energy state the electronic state gets.
Therefore, the lattice is sometimes largely distorted, up to 10\% of a lattice constant, as $1T$-TaSe$_2$\cite{tase2}.
In the ordered CDW phase, the electronic states produce a large gap between the bonding and anti-bonding states\cite{suzukiharima},
where the strong Coulomb repulsion does not play an important role.
On the contrary, when the Coulomb interaction plays a crucial role, the distortion is kept very small; e.g. only about 0.05\% in PrRu$_4$P$_{12}$\cite{lee},
where the lattice distortion itself does not need to gain the lower electronic states\cite{harima}.
The anti-ferro total-symmetric multipole ordering breaks the translational symmetry to open the gap in the conduction band, then the lattice is slightly, and symmetrically distorted in order to relax multipoles developed on Pr-sites.

URu$_2$Si$_2$ shows a large peak in specific heat at 17.5 K ($T_{\rm O}$), indicating a second order phase transition, then the ordered state coexists with superconductivity found below 1.5 K\cite{1985,1985b,1985c}.
Although the intensive theoretical and experimental studies have been performed, the order parameter of the phase below $T_{\rm O}$ is still unknown,
so now it is called the hidden order (HO).
Quite different theoretical proposals have appeared\cite{santini,barzykin,ikeda,ohkawa,chandra,kiss,varma,balatsky,8}:
some neglecting the itinerant character of the $5f$ electrons, 
and some by omitting the local specificity of the U site;
in the U$^{4+}$ valence state ($5f^{2}$ configuration), multipole ordering can be expected as recently discussed,
in the physics of Pr skutterudite systems\cite{skutterudite}.
 
On the experimental side, it seems now well established that the low pressure HO phase is not antiferromagnetic, 
but that above $P_x \sim 0.5$ GPa through the first order transition long range anti-ferromagnetism (AF) is established with wave vector $\mathbf Q_{AF}=(0, 0, 1)$ corresponding to a lattice doubling along the $c$ axis\cite{9,10,Flouquet}.
Furthermore, the HO phase is characterized by a sharp resonance at energy $E_0=1.5$ meV observed at the wave vector $\mathbf Q_{\rm O}=(1, 0, 0)$ (equivalent to $(0, 0, 1)$ in body center tetragonal lattice). 
Switching to the AF phase at $P_x$ leads to the collapse of the resonance and to the observation of elastic neutron magnetic signal at $\mathbf Q_{\rm O}$\cite{Flouquet}. 
Taking into account that the same characteristic vector $\mathbf Q_{\rm O}$ emerges in the HO and the AF phase as well as the invariance of three Shubnikov de Haas frequencies through $P_x$\cite{12}, 
the experimental data push to look to a model, 
where the transition from PM phase to HO or AF ground states corresponds to the same structural transition among the tetragonal classes,
while the pressure switching from HO to AF adds a supplementary time reversal breaking. 

A previous band structure calculation demonstrates that AF opens a large gap at the Fermi level,
then the proposal was that the occurrence of large magnetic fluctuations may preserve a Fermi surface construction at $T_{\rm O}$ in the HO phase similar to that proposed going from PM to AF above $P_x$\cite{8}
Recently, by using LDA+DMFT method, an idea quite similar to our proposal has been developed by Kotliar and coworkers\cite{13},
but the selected order parameter of HO is a hexadecapole.

In this letter, we propose that the space group in the HO phase is No.~136, then the transition at 17.5 K does not need any kind of lattice distortion, 
and keeps Ru-site with 4-fold symmetry.
Therefore, most of the experimental techniques are unable to detect the characteristic charge distribution in the HO phase.
This is one simple answer to why the HO is still hidden.

%\section*{Space group analysis}

\begin{table*}[htb]%[htbp]
\caption{Multiplicity, Wyckoff letter, and site symmetry of each atom in URu$_2$Si$_2$. First row; the case of No.~139, the mother space group: Below the second row; the maximal non-isomorphic $k$ subgroups of No.~139.
All the subgroups belong to $D_{4h}$.
The site coordinates are shown, when the notation changes.
In No.~131, the origin of the lattice changes, but the lattice does not need to be distorted.
}
\label{t1}
\begin{center}
\begin{scriptsize}
\begin{tabular}{llll}
\hline
\multicolumn{1}{c}{Space group} & \multicolumn{1}{c}{U-site} & \multicolumn{1}{c}{Ru-site} & \multicolumn{1}{c}{Si-site} \\
\hline
No.~139 $ I4/mmm$ & $2a\ 4/mmm\ (0, 0, 0)$ & $4d\ \bar{4}m2\ (0, {1\over 2}, {1 \over 4}), ({1\over 2}, 0, {1 \over 4})$ & $4e\ 4mm\ (0, 0, z), (0, 0, \bar{z})$ \\
\hline
No.~123 $ P4/mmm$ & $1a\ 4/mmm$  \textcircled{+}\ $1d\ 4/mmm$ & $4i\ 2mm.\ (0, {1\over 2}, z), ...$ & $2g\ 4mm\ (0, 0, z), ...$   \textcircled{+}\ $2h\ 4mm\  ({1\over 2}, {1\over 2}, z'), ...$\\
%& \textcircled{+}\ $1d\ 4/mmm $ && \textcircled{+}\ $2h\ 4mm\  ({1\over 2}, {1\over 2}, z'), ...$ \\
No.~126 $ P4/nnc$ & $2a\ 422$ & $4d\ \bar{4}.. $ & $4e\ 4..$ \\
No.~128 $ P4/mnc$ & $2a\ 4/m $ & $4d\ 2.22 $ & $4e\ 4.. $ \\
No.~129 $ P4/nmm$ & $2c\ 4mm\ (0, {1\over 2}, z), ...$ & $2a\ \bar{4}m2\ (0, 0, 0), ...$ \textcircled{+}\ $2b\ \bar{4}m2\ (0, 0, {1\over 2}), ... $ & $2c\ 4mm\ (0, {1\over 2}, z), ...$ \textcircled{+}\ $2c\ 4mm\ (0, {1\over 2}, z'), ... $\\
%& & \textcircled{+}\ $2b\ \bar{4}m2\ (0, 0, {1\over 2}), ... $ & \textcircled{+}\ $2c\ 4mm\ (0, {1\over 2}, z'), ... $ \\
No.~131 $ P4_2/mmc$ & $2c\ mmm.\ (0, {1\over 2}, 0), ...$ & $2e\ \bar{4}m2\ (0, 0, {1 \over 4}), ...$ \textcircled{+}\  $2f\ \bar{4}m2\ ({1\over 2}, {1\over 2}, {1 \over 4}), ...$ & $4i\ 2mm.\  (0, {1\over 2}, z), ...$ \\
%& & \textcircled{+}\  $2f\ \bar{4}m2\ ({1\over 2}, {1\over 2}, {1 \over 4}), ...$ & \\
No.~134 $ P4_2/nnm$ & $2a\ \bar{4}2m$ & $4d\ 2.22$ & $4g\ 2.mm$ \\
No.~136 $ P4_2/mnm$ & $2a\ m.mm $ & $4d\ \bar{4}..$ & $4e\ 2.mm$ \\
No.~137 $ P4_2/nmc$ & $2a\ \bar{4}m2$ & $4d\ 2mm.\ (0, {1\over 2}, z), ...$ & $4c\ 2mm.$ \\
\hline
\end{tabular}
\end{scriptsize}
\end{center}
\end{table*}

After a second order phase transition, the lower symmetry must be one of the subgroups of the mother symmetry.
The space group in high temperature phase of URu$_2$Si$_2$ (ThCr$_2$Si$_2$-type) is No.~139 ($I4/mmm$; $D_{4h}^{17}$), 
which has 7 maximal non-isomorphic $t$ subgroups ({\it translationengleiche}), 
8  maximal non-isomorphic $k$ subgroups ({\it klassengleiche}), and two maximal isomorphic subgroups of lower index\cite{Table}.
$t$ subgroups
(No.~69 ($Fmmm$; $D_{2h}^{23}$), 
No.~71 ($Immm$; $D_{2h}^{25}$), 
No.~87 ($I4/m$; $C_{4h}^5$), 
No.~97 ($I422$; $D_4^9$),
No.~107 ($I4mm$; $C_{4v}^9$), 
No.~119 ($I\bar{4}m2$; $D_{2d}^9$), 
No.~121 ($I\bar{4}2m$; $D_{2d}^{11}$)) change the point group, and the isomorphic subgroup needs a super structure (($c'=3c$) or ($a'=3a$, $b'=3b$)), 
so they should be easily observed by experiments.
As the crystal class is unchanged, the point group of 8  maximal non-isomorphic $k$ subgroups must be considered.

Multiplicity, Wyckoff letter, and site symmetry of each atom in URu$_2$Si$_2$ are listed in Table~\ref{t1}, 
with all the $k$ subgroups.
All the space groups have the point symmetry $D_{4h}$.
(from No.~123 ($D_{4h}^1$) to No.~142 ($D_{4h}^{20}$))
Let us now investigate the correspondence of atom positions between the mother group and the $k$ subgroups. 
No.~139 ($I4/mmm$; $D_{4h}^{17}$) belongs to the body centered tetragonal lattice, then 16 point symmetry operations and $(\frac{1}{2}, \frac{1}{2}, \frac{1}{2})$ translation in the conventional lattice.
Totally 32 symmetry operations transfer an atom to the crystallographically equivalent position, or rotate/reflect an atom at the same position.
For example, U at (0, 0, 0) in No.~139, is transferred by $(\frac{1}{2}, \frac{1}{2}, \frac{1}{2})$, but not by  the 16 point symmetry operations, 
so that the order of the site symmetry ($4/mmm$; $D_{4h}$) is 16.
The multiplicity of the site (2) times the order of the site symmetry (16) is 32.
In the case of Ru and Si, the multiplicity of the site is 4, and the order of the site symmetry is 8 (($\bar{4}m2$; $D_{2d})$ or ($4mm$; $C_{4v}$)), as listed in Table~\ref{t1}.

In the maximal subgroups, 16 symmetry operations remain.
The lost operation brings the splitting of the site into two inequivalent sites, or the lowering of the site symmetry.
When the site is split as U and Si in No.~123, Ru and Si in No.~129, or Ru in No.~131, conventional experimental techniques can detect it.
Even when the site is not split, the internal parameter arises in $4i$ of No.~123, $2c$ of No.~129, and $4d$ of No.~137, resulting in atoms moving.
However, any kind of lattice distortion is not expected, when the lower space group is No.~126, No.~128, No.~134, or No.~136. 

NQR measurement is often a powerful tool to probe the phase transition\cite{mito,kouchi}.
Ru-site NQR measurements have been already preformed\cite{takagi}, but never show any anomaly around at $T_{\rm O}$.
If the Ru-site loses the 4-fold symmetry axis due to the phase transition, 
asymmetric parameter $\eta$ becomes non zero to affect the NQR frequency.
The unchanged NQR frequency indicates that the 4-fold axis of Ru-site survives even in lower temperatures.
Therefore, the possibility of No.~128 or No.~134 is eliminated.

\begin{figure}[tb]%[htbp]
\begin{center}
\includegraphics[width=1.0\linewidth]{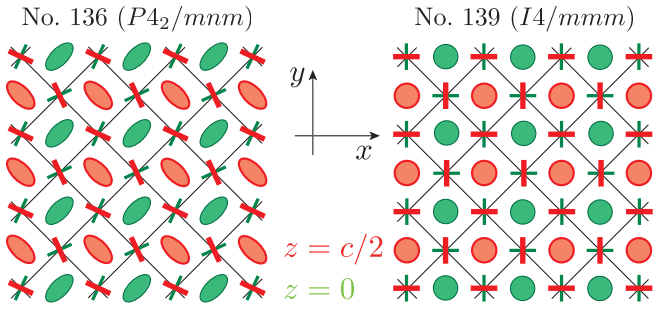}
\caption{Left: Schematic charge density map of the HO phase. 
Green and red ovals represent charge densities of U-sites. 
Green (red) sites situate on the $z=0$ ( $z=c/2$) plane. 
This is the anti-ferro $O_{xy}$ type quadrupole ordering. 
The crosses of green and red bars represent the schematic charge distribution at Ru sites on the $z=c/4$ plane, 
which keeps the $S_4$ symmetry even in the ordered phase. 
%Note that the NQR frequency never show any anomaly at the transition temperature. 
The charge distribution of Si (not shown) also lost the 4-fold symmetry, just like U-site. 
Right: Schematic charge density map of the disordered phase. 
From right to left, any kind of lattice distortion is not allowed.}
\label{order136}
\end{center}
\end{figure}

In No.~126 the local inversion symmetry at U-site is lost, then the U site belongs to $422$; $D_4$,
while in No.~136 the local $C_4$ ($\pi/2$ rotation) at U-site is lost, then the U site belongs to $mmm$; $D_{2h}$.
In both cases, not the lattice distortion, but the charge distortion at the U-site, caused by large Coulomb interaction between $5f$ electrons, must be the origin of the symmetry lowering.
The charge distortion without inversion symmetry, could not be happened, without moving atoms.
However, the charge distortion without $C_4$ could be possible --- it is ascribed to quadrupole ordering, 
i.e., a second order phase transition from No.~139 to No~136 could be compatible with $O_{xy}$-type anti-ferro quadrupole ordering with $\mathbf Q=(0, 0, 1)$,
which does not couple with any lattice distortion.
The schematic charge density map of the HO phase is shown in  Fig.~\ref{order136}.
$\mathbf Q=(0, 0, 1)$ is equivalent to $\mathbf Q=(1, 0, 0)$ or $(0, 1, 0)$ in the left panel of Fig.~\ref{order136}.

It is a naive consequence that the space group in the HO phase is No.~136.  
To conclude it, it is not necessary to claim the $5f^2$ local picture for URu$_2$Si$_2$.
However, we take the local picture in the following discussion, because it is more easy to discuss the origin of the transition.

%\section{What lowers the energy in the hidden ordered phase}

As mentioned in the introduction, the low temperature phase should have a lower energy.
Now, we propose the phase transition which does not require any movement of atoms.
Therefore, the mechanism to reach the lower energy must be based on large Coulomb interactions.
However, the anti-ferro quadrupole ordering itself does not lower the system energy,
but raises it due to the local electronic charge distortion.
We need a strong coupling constant for the anti-ferro quadrupole interaction.

One may expect nesting character in the Fermi surface of the non-$f$ reference compound ThRu$_2$Si$_2$,
as in LaRu$_4$P$_{12}$ for PrRu$_4$P$_{12}$.
In the local $4f^{2}$ picture of PrRu$_4$P$_{12}$, corresponding to LaRu$_4$P$_{12}$,
the conduction band with strong nesting character provides the large coupling constant for the higher multipole interaction\cite{harima}.
However, no clear indication of such nesting property observed in the band structure of ThRu$_2$Si$_2$, 
 as shown later.

In this letter, we examine which kind of orbitals can gain energy by  introducing anti-ferro quadrupole ordering.
For the simplicity, we take one dimensional chain along the $z$-direction with $p_x$ and $p_y$ orbitals at each site.
The lattice constant is $a$, and each lattice point have 2 sites with a distance $a/2$.
Therefore, the energy dispersion is expressed by $E=\ve +2t\cos(k{{a}/2})$ for $|k|  \le {2\pi/a}$ for each $p_x$ and $p_y$ orbital,
where $\ve$ is the origin of the $p_x$ and $p_y$ orbital, $\ve_x$ and $\ve_y$.
$t$ ($< 0$) is the transfer or two center integral  ($(pp\pi)$) between $p_x$ or $p_y$ orbitals along the $z$-direction.
Then, let us now take a unit cell with the lattice constant $a$, (then $|k|  \le {\pi/a}$)
which contain two sites, $A$ and $B$ with $p_x$ and $p_y$ orbitals each.
The energy eigenvalues are expressed by the matrix;

\begin{equation}
\left(\begin{matrix}
{\hskip9mm \ve_x^A\hskip5mm , 2t\cos ({{k a}/ 2}), \hskip8mm 0 \hskip7mm, \hskip8mm 0 \hskip7mm}\\
{2t\cos ({{k a}/ 2}), \hskip9mm \ve_x^B\hskip5mm , \hskip8mm 0 \hskip7mm, \hskip8mm 0 \hskip7mm}\\
{\hskip8mm 0 \hskip7mm, \hskip8mm 0 \hskip7mm, \hskip9mm \ve_y^A\hskip5mm , 2t\cos ({{k a}/ 2})}\\
{\hskip8mm 0 \hskip7mm, \hskip8mm 0 \hskip7mm, 2t\cos ({{k a}/ 2}), \hskip9mm \ve_y^B\hskip5mm }
\end{matrix}\right),
\end{equation}
where the $\ve_x^A$ is the origin of the $p_x$ orbital at $A$-sublattice site.
Now obviously $\ve_x^A=\ve_x^B=\ve_y^A=\ve_y^B$, so the 4 states are degenerated at the zone boundary $|k|={\pi/a}$.

Once 
we introduce an anti-ferro quadrupole ordering from $A$ sublattice site to $B$ sublattice site, 
$\ve_x^A=\ve_y^B\ne \ve_x^B=\ve_y^A$, resulting in a gap opening at the zone boundary $|k|={\pi/a}$.
The similar scenario has been discussed for filled skutterudites PrRu$_4$P$_{12}$,
where a total symmetric anti-ferro multipole ordering is introduced\cite{harima}.
Here the gap ($|\ve_x^A- \ve_x^B|$) originates in the $O_{xy}$-type $5f^2$ charge distortion.
This situation in band electrons with a gap at the zone boundary is very easily understood, 
when one imagines an anti-ferro spin ordering, then replaces the up and down spins with $p_x$ and $p_y$ orbitals.
Obviously, the energy gain becomes the maximum when the Fermi level is lying in the gap.
But, the energy is more or less lowered, if the bands are not fully occupied.

\begin{figure}[tb]%[htbp]
\begin{center}
\includegraphics[width=0.9\linewidth]{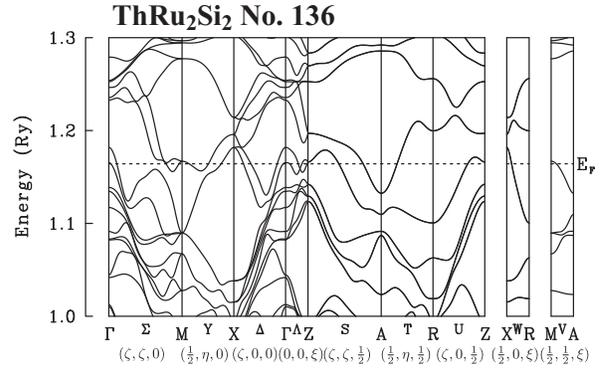}
\caption{The electronic band structure of ThRu$_2$Si$_2$ within the space group No.~136.
A self-consistent calculation results in invisible anti-ferro $O_{xy}$-type potential, 
so this is the same as the band structure that is calculated within the space group No.~139 (body center tetragonal), 
then folded into the 1st Brillouin Zone of No.~136 (simple tetragonal).
%The spin-orbit interactions are included.
Note that two bands are degenerated on the boundary of the Brillouin Zone (the Y, S, T, U, W, and V axes).
Among them the degeneracy along the Y and V axes is due to the symmetry under No.~136;
while the S, T, U and W axes are accidental, so they will split once anti-ferro $O_{xy}$-type potential is introduced.
No flat band dispersion appears near the Fermi level, E$_{\rm F}$, along the zone boundary axes; the S, T, and U axes on $ |k_z|=\frac{1}{2}(\frac{2\pi}{c})$ plane.
}
\label{band}
\end{center}
\end{figure}

In the real three dimensional system, flat band dispersion on the zone boundary $|k_z|=\pi/c$ is necessary to realize a clear gap opened.
However, such flat dispersion could not be observed in the band structure of ThRu$_2$Si$_2$, which is shown in Fig.~\ref{band}.
Even unless it is a clear gap opened in the vicinity of the Fermi level, 
introducing  anti-ferro $O_{xy}$-type potential can lower the energy of the itinerant electrons with $\{x, y\}$- or $\{xz, yz\}$-type component.
%In the band structure of ThRu$_2$Si$_2$ in No.~136 as shown in Fig.~\ref{band}, 
%the accidental degeneracy remains on the zone boundary.

%because anti-ferro $O_{xy}$ type ordering has not yet appeared spontaneously.
The band structure of ThRu$_2$Si$_2$ in Fig.~\ref{band} is calculated by assuming No.~136 space group, but using the same lattice as for the No.139.
The spin-orbit interaction is included , so no degenerate bands appear even at the $\Gamma$ point.
(E$_{{\rm g}}$: $\{xz, yz\}$ or E$_{{\rm u}}$: $\{x, y\}$, unless the spin-orbit interaction)
The bands in the vicinity of the Fermi level, consist of Ru-$d$ and Th-$d$ components. 
The anti-ferro $O_{xy}$-type potential yields the matrix element between $\{x, y\}$- or $\{xz, yz\}$-type orbitals of them, 
then the system energy would be lowered.
%It should be noted that even in a conventional LDA calculation for URu$_2$Si$_2$ in No.~136, the orbital could not be ordered.
This scenario is just a speculation so far.
However, once it is ordered, the conduction bands are strongly modified, 
then  the carrier number would be reduced as observed in the HO phase\cite{behnia1,kasahara}.
A realistic band structure calculation of the HO phase is now in progress by using LDA+$U$ method.

The Coulomb interaction between the local $5f$ quadrupole moment pair ($O_{\pm xy}^A$-$O_{\mp xy}^B$) and the conduction band pair ($(x\pm y)^A$-$(x\mp y)^B$) can be written as $-I (S\cdot s)$, where $S$ ($s$) is an Ising type spin representing $+$ or $-$ for quadrupole states (orbital states).
If it is written as ($\mathbf S\cdot \mathbf s$), the Kondo behavior could be expected, as discussed below.
It is emphasized that the coupling $I$ is independent from lattice distortions in this system.
This coupling survives even in high temperatures escaping from disturbance of the lattice vibrations.

Now, it is worth mentioning the CEF splitting of $5f^2$ electrons in the local picture.
It is not necessary to take a ground state degenerated.
Due to the change of local symmetry from $D_{4h}$ to $D_{2h}$, 
off-diagonal term may emerge to lower the ground state, even if it is a singlet.
Then the magnetic susceptibility shows a cusp at $T_{\rm O}$, as observed.\cite{amiami}.

%\section*{CEF Scheme and Physical Properties}
Anti-ferro quadrupole ordering has been discussed before, by using three singlet states\cite{santini} or a doublet ground state\cite{ohkawa}.
They did not distinguish between $O_{xy}$-type and $O_{x^2-y^2}$-type, but taking $O_{xy}$-type and forming anti-ferro coupling are especially important.
In this paper, we will discuss shortly that
the present HO state can be understood in a manifold of one singlet state, 
$\Gamma_{1}$ (A$_{1{\rm g}}$; $z^{2}$), and one doublet state, 
$\Gamma_{5}^{\pm}$ (E$_{{\rm g}}$; $xz(+)$, $yz(-)$).

Eigen functions of these states are given as 
$|\Gamma_{1}\rangle=\varepsilon(|4\rangle+|-4\rangle)+\gamma|0\rangle$, 
and $|\Gamma_{5}^{\pm}\rangle=\alpha|\pm 3\rangle+\beta|\mp1\rangle$~\cite{santini}.   
There exist non-vanishing matrix elements of 
$Q_{xz}\equiv \{J_{x},J_{z}\}/2$ and $Q_{yz}\equiv \{J_{y},J_{z}\}/2$ 
between $\Gamma_{1}$ and $\Gamma_{5}^{\pm}$: i.e., 
$\langle\Gamma_{1}|Q_{xz}|\Gamma_{5}^{\pm}\rangle=\pm c_{{\rm Q}}$, and 
$\langle\Gamma_{1}|Q_{yz}|\Gamma_{5}^{\pm}\rangle=\mp{\rm i}c_{{\rm Q}}$, 
with $c_{\rm Q}\equiv(3\sqrt{2}\alpha\varepsilon-\sqrt{5}\beta\gamma)$.  
The HO of No.136 can be identified as an induced-moment AF order of 
$Q_{xz}+Q_{yz}$ at $A$ sublattice points and $Q_{xz}-Q_{yz}$ 
at $B$ sublattice points, respectively.  
%as in the case of moment-induced magnetic 
%order in CEF singlet systems such as Pr-based compounds~\cite{Grover}.  
Note that the symmetry of charge distribution is $O_{xy}$: 
$[z(x\pm y)]^{2}=z^{2}(x^{2}+y^{2}\pm2xy)$.  
In the Hilbert space spanned by $\Gamma_{1}$ and $\Gamma_{5}^{\pm}$, the 
quadrupole operator ${\hat Q}_{xz}$ and ${\hat Q}_{yz}$ are represented as 
\begin{eqnarray}
{\hat Q}_{xz}&=&c_{\rm Q}[(|\Gamma_{1}\rangle\langle\Gamma_{5}^{+}|
+|\Gamma_{5}^{+}\rangle\langle\Gamma_{1}|)-
(|\Gamma_{1}\rangle\langle\Gamma_{5}^{-}|
+|\Gamma_{5}^{-}\rangle\langle\Gamma_{1}|)],
\hskip7mm\label{Q1}
\\
{\hat Q}_{yz}&=&-{\rm i}c_{\rm Q}[(|\Gamma_{1}\rangle\langle\Gamma_{5}^{+}|
-|\Gamma_{5}^{+}\rangle\langle\Gamma_{1}|)-
(|\Gamma_{1}\rangle\langle\Gamma_{5}^{-}|
-|\Gamma_{5}^{-}\rangle\langle\Gamma_{1}|)].
\hfill\label{Q2}
\end{eqnarray}
These two operators form the SU(2) Lie algebra together with their comutator.  Corresponding to the quadrupole operator ${\hat Q}_{xz}$ and 
${\hat Q}_{yz}$ in the $f^{2}$-configuration, we can define the quadrupole 
operators ${\hat q}_{xz}$ and ${\hat q}_{yz}$in the $f^{1}$-configuration. 
Then, from a symmetry argument, there exists a Kondo type interaction 
\begin{equation}
 H_{\rm K}=J_{\rm Q}({\hat Q}_{xz}{\hat q}_{xz}+{\hat Q}_{yz}{\hat q}_{yz}). 
\label{Q7}
\end{equation}
where ${\hat Q}_{xz}$ (${\hat q}_{xz}$) and ${\hat Q}_{yz}$ (${\hat q}_{yz}$) 
are quadrupole coordinates of U-ion with $f^{2}$-configuration 
(of conduction electrons described by the symmetry of $f^{1}$-configurations).  
This interaction can give rise to an ``induced-moment Kondo effect" as in the case of 
Pr-based skutterudite compounds~\cite{Hattori}.  Detailed discussions will be 
given in a separate paper.  

A Nambu-Goldstone (NG) mode around this HO state is expected to consist of 
doubly degenerate excited states $\Gamma_{5}^{\pm}$ and the singlet 
ground state $\Gamma_{1}$, as in the case of ``gapless spin waves" in CEF 
singlet systems such as Pr-based compounds~\cite{Grover}.  
%From an intuitive consideration, 
%the gapless NG mode should be linear combinations $z(x-y)$ at A　sub　
%lattice site and $z(x+y)$ at B　sub　lattice site, because this fluctuation 
%mode recover the symmetry in the $xy$-plane.  
Quite recently, two characteristic phenomena which seem to have a potentiality to 
unveil the form of HO were reported. One is resistivity measurements at low 
temperature region: the resistivity along the $a$-axis, $\rho_a$, exhibits a 
non-Fermi liquid behavior, $\rho_{a}=\rho_{0a}$+$A_{a}T$, and that along the 
$c$-axis shows the usual Fermi liquid behavior, 
$\rho_{c}=\rho_{0c}$+$A_{c}T^{2}$\cite{Behnia}. 
Another one is neutron scattering measurement: it was reported on neutron 
scattering experiment at $P=0.67$ GPa that an inelastic mode of magnetic 
fluctuations $m_{z}$ at the wave vector at $\mathbf Q=(1,0,0)$ exists only 
in the HO phase, not in the normal nor the large moment antiferromagnetic 
phases.\cite{10}
These two unusual phenomena can be explained 
if the NG mode in HO state is a linear combination of quadrupolar 
fluctuation of $Q_{xz}$ and $Q_{yz}$ representing $\Gamma_{5}^{\pm}$ state in 
tetragonal symmetry\cite{Miyake1}.
The detailed analysis is now in progress.

In higher pressure phase, large moment antiferromagnetism (LMAF) of Ising type 
(with $c$-axis as an easy axis) is realized.  In order to treat the LMAF and 
HO order on an equal footing, we have to extend the Hilbert space, 
spanned by $|\Gamma_{1}\rangle$ and $|\Gamma_{5}^{\pm}\rangle$, by adding 
the CEF state $\Gamma_{2}$ (A$_{2{\rm g}}$; $xy(x^{2}-y^{2})$) whose eigen 
function is $|\Gamma_{2}\rangle=(|4\rangle-|-4\rangle)/\sqrt{2}$.  
The detailed  discussions are discussed in a separate paper.

The key point remains to observe the symmetry lowered in HO.
As many unsuccessful attempts have been tried; obviously improvements in the method's sensitivity are necessary. 
For example careful check of the anisotropy of the magnetization in the basal plane above $T_{\rm O}$  will be worthwhile 
as short correlations of the quadrupole pair ($O_{xy}$-$O_{-xy}$) must enhance a 4 fold component. 
Refinement in resonant X ray measurements (well-known tool in the proof of quadrupole ordering) is required as the last result fails to detect any quadrupole order.\cite{rxray}
Further experimental efforts are greatly encouraged.

%\section{Conclusion}
In conclusion, the second order transition to the HO phase in URu$_2$Si$_2$ does not need any type of lattice distortions.
URu$_2$Si$_2$ is never a unique material without structural distortion, 
but the heart of materials in the strongly correlated electron systems, where only electrons themselves bring about lowering spatial symmetry.

\section*{Acknowledgment}
We thank H. Amitsuka, K. Behnia, Y. Matsuda, T. Sakakibara, J.A. Mydosh, A, Kiss, C. Pepin, T. Mito and Z. Hiroi for useful and fruitful comments.
We are pleased to dedicate this paper to Patrik Fazekas, for his great service to this field.
HH also thanks Aspen Center for Physics for his two week stay on the related subject.
%and his daughters for staying six weeks in Grenoble together with him.
This work was partly supported by a Grant-in-Aid for Scientific Research on Innovative Areas "Heavy Electrons" (No. 20102002) of The Ministry of Education, Culture, Sports, Science, and Technology, Japan.

\end{document}